\documentclass[twocolumn, prl]{revtex4}

\usepackage{graphicx}
\usepackage{dcolumn}
\usepackage{bm}

\begin{document}

\title{ A Matrix Kato-Bloch Perturbation Method for Hamiltonian 
Systems}

\author {S. Moukouri}

\affiliation{ Michigan Center for Theoretical Physics and
             Department of Physics, \\
         University of Michigan 2477 Randall Laboratory, Ann Arbor MI 48109}

\begin{abstract}
A generalized version of the Kato-Bloch perturbation expansion is presented.
It consists of replacing simple numbers appearing in the perturbative
series by matrices.
This leads to the fact that the dependence of the eigenvalues of the 
perturbed system on the strength of the perturbation is not necessarily
polynomial. The efficiency of the matrix expansion is illustrated in three 
cases: the Mathieu equation, the anharmonic oscillator and weakly coupled 
Heisenberg chains. It is shown that the matrix expansion converges for
a suitably chosen subspace and, for weakly coupled Heisenberg chains, it 
can lead to an ordered state starting from a disordered single chain. This 
test is usually failed by conventional perturbative approaches.     
\end{abstract}

\maketitle

 Since its introduction long ago by Rayleigh and Schr\"odinger, the perturbation
method has become an essential tool for the analysis of Hamiltonians.
  The perturbative method consists in searching for 
an approximate solution of the eigenvalue equation of a linear operator 
slightly different from an unperturbed operator whose spectrum is known. 
In the perturbation expansion it is assumed that, if the system Hamiltonian is 
written as the sum of two parts $H=H_0+gV$, where $H_0$ is the unperturbed
Hamiltonian and $gV$ the perturbation, the perturbed
energies and eigenfunctions are power series of $g$.  The convergence of these
 series was studied by mathematicians \cite{kato_BOOK}, whose work may be 
summarized by two important results.

 Let $A$ be a unitary space (i.e., a Hilbert space or more generally 
a Banach space). The eigenvalues of $H$ satisfy the characteristic equation 
$det(H-\lambda)=0$. This is an algebraic equation of degree $N$ equal to
the dimension of $A$, $N=dimA$. Its coefficients are holomorphic
functions of $g$. The first result is given by the following theorem:
  {\it the roots $\lambda$ of $det(H-\lambda)=0$  constitute one
or several branches of analytic functions of $g$ with only algebraic
singularities}. The second result is a theorem due to Rellich \cite{kato_BOOK}, 
 and may be stated as follows: {\it the power series for an eigenstate or an
eigenvalue of $H$ are convergent if the magnitude of the perturbation
$|gV|$ is smaller than half of the isolation distance of the corresponding
unperturbed eigenvalue}. In other terms, if an eigenvalue of $H_0$ is isolated
 from the rest of the spectrum then, for a sufficiently small $g$,
the expansion will converge.

These results have been known for a long period of time. But they
are barely mentioned in numerous works in condensed matter theory 
or in high energy theory in which perturbation expansions are used. 
This is because, in most problems, one is mainly interested in one or 
two-particle properties. Hence, perturbation expansions were formulated 
in terms of single or two-particle Green's functions of the unperturbed 
Hamiltonian \cite{nozieres}. The convergence properties of Green's function
is very tedious to analyze.  One has to sum up complicated classes of 
infinite series of Feyman diagrams, in the so-called parquet summation,  
without a clear selection criterion. This type of expansion generally
leads to the divergence of the dominant quantum fluctuations at
low temperatures. This indicates the onset of long-range order, but,
the ordered state cannot be reached. Reaching the ordered state may 
necessitate going across energy levels or a singularity. This is impossible with
the conventional perturbation expansion which assumes the ground-state
energy to be an analytic function of $g$. This has lead to the widespread
belief that the spectrum of $H_0$ (which is supposed to describe 
a disordered phase) is not a good starting point if one wishes to reach the 
ordered phase. But a clear mathematical criterion on the convergence of
the Green's function series has never been provided. Hence, the 
Green's function method simplifies the mathematics of the problem by 
reducing the many-body problem to a one- or two-body problem, but, it blurs
the analysis of eventual convergence problems.

It could thus be better to analyze the perturbative expansions of Hamiltonians
in the light of the above two theorems by using stationary perturbation
instead of Green's functions. However it is clear that if $H_0$ is the 
hopping Hamiltonian as used in Fermi systems, the Rellich theorem does
not apply. This is because the spectrum is gapless and thus any 
perturbation, no matter how small, will not fullfil the convergence 
condition. That is why it is crucial to work in a finite volume. But, even 
in this case, most perturbation expansions do not converge if the system 
is large for reasonable values of $g$.

In this letter, a simple cure for the convergence problems of perturbative 
expansions defined in a finite volume is proposed. The starting point is 
the general perturbative expansion derived by Kato \cite{kato} and 
Bloch \cite{bloch}. The new method consists in replacing the original 
Kato-Block simple polynomial series  by a matrix expansion. In the matrix
expansion, the low lying excited states are used to shield the ground state
from the rest of the spectrum so that the near degeneracy problems that
usually plague conventional perturbation expansions are avoided. Arguments, but
not a mathematical demonstration, of the convergence of the matrix expansion
are given. The method is tested in some simple cases including the Mathieu
equation and the anharmonic oscillator. Then, the new method is applied to
 weakly coupled antiferromagnetic (AFM) chains, a problem of high current 
interest in the physics of low-dimensional magnetic materials 
\cite{scalapino,schulz}. It is shown that, starting from a 
disordered ground state of a single chain, the new matrix perturbation is 
able to reach the ordered state when a small exchange interaction 
is turned on between the chains.

 The Kato-Bloch expansion, for the correction ${\tilde E}_0$
to an eigenset ($E_0$, $|\phi_0 \rangle$), which is supposed to be non-degenerate, 
is given by \cite{messiah}:

\begin{eqnarray}
P_0{\tilde E}_0P_0=P_0H_0P_0 + gP_0VP_0+g^2P_0V{\tilde Q}_0VP_0+...
\label{expansion0}
\end{eqnarray}

\noindent where $P_0=|\phi_0 \rangle \langle \phi_0|$, and 

\begin{eqnarray}
{\tilde Q}_0=\sum_{n>0} \frac{|\phi_n \rangle \langle\phi_n|}
              {E_0-E_{n}}.
\label{projector0}
\end{eqnarray}

When the problem is projected onto the subspace generated by the
eigenstate $\phi_0$, one retrieves the perturbative series

\begin{eqnarray}
{\tilde E}_0=E_0+g\langle \phi_0|V|\phi_0 \rangle+g^2 \sum_{n >0} 
\frac{|\langle \phi_0|V|\phi_{n} \rangle|^2}{E_0-E_{n}}+...
\label{expansion1}
\end{eqnarray}

\begin{figure}
\includegraphics[width=3. in]{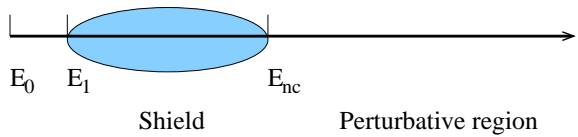}
\caption{Sketch of the separation of the spectrum in two regions.}
\label{shield}
\end{figure}

In the expansion of (~\ref{expansion1}), higher order correction terms to
the ground state energy ${\tilde E}_0$ are governed by the ratio 
$g/(E_{1}-E_0)$. Kato showed that if $g/(E_1-E_0) < 1/2$, then,
the series converges. The problem that may lead to convergence failure of 
the perturbative series is easy to see in the Kato-Bloch formulation: the 
series diverges if  $g/(E_1-E_0) > 1/2$. When this occurs, it is possible
to eliminate the problem by formulating the expansion (~\ref{expansion0}) 
in terms of matrices. This can lead to  convergence for a suitable size of the 
matrices. The idea behind the matrix expansion is to shield $E_0$ from the 
rest of the spectrum in the expansion (Fig.~\ref{shield}). i.e., 
instead of restricting
$P_0$ to the eigenvalue of interest $E_0$, one also includes a few excited
states above $E_0$ up to the cut-off energy $E_{n_c}$. In the matrix
method, $P_0$ is now given by

\begin{eqnarray}
P_0=\sum_{n}^{n_c} |\phi_{n}\rangle \langle \phi_{n}|,
\label{projector1}
\end{eqnarray} 

\noindent and the complement of the projector which enters into the perturbation
expansion is

\begin{eqnarray}
{\tilde Q}_0=\sum_{n>n_c} \frac{|\phi_{n}\rangle \langle \phi_{n}|}
              {E_0-E_{n}}.
\label{projector2}
\end{eqnarray}

In this case, when the problem is projected onto
the subspace generated by the $\phi_{n}$, $n=0,...,n_c$, 
each term in the expansion of the corrected energy of (~\ref{expansion1}) 
is replaced by a matrix. Now, the largest term of order $k$ in the matrix 
expansion is $g^k/(E_{n_c+1}-E_0)^{k-1}$ instead of $g^k /(E_1-E_0)^{k-1}$ 
so that the condition $g/(E_{n_c+1}-E_0) < 1/2$ is fulfilled for a suitable
chosen $n_c$.

 If the matrix method is used with truncation to two states $E_0$ and $E_1$,
the Kato-Block matrix expansion leads to the Hamiltonian ${\tilde H_0}$
given by 

\begin{eqnarray}
\nonumber {\tilde H}_0 \approx \left( 
\begin{array}{cc} 
E_0 & 0 \\
0 & E_1 
\end{array}
\right)+g
 \left( 
\begin{array}{cc} 
V_{00} & V_{01} \\
V_{10}     & V_{11} 
\end{array}
\right)- \\
g^2
\left( 
\begin{array}{cc} 
V_{00}^{(2)} & V_{01}^{(2)} \\
V_{10}^{(2)} & V_{11}^{(2)} 
\end{array}
\right)+...,
\label{block1}
\end{eqnarray}

\noindent where the second order matrix elements projected to the two 
states kept in the matrix expansion are respectively

\begin{eqnarray}
V_{00}^{(2)}=\sum_{n>1}\frac{V_{0n}^2}{E_n-E_0}, \\
V_{01}^{(2)}=\sum_{n>1}\frac{V_{0n}V_{1n}}{E_n-E_0},\\
V_{11}^{(2)}=\sum_{n>1}\frac{V_{1n}^2}{E_n-E_0}.
\end{eqnarray}

 In the matrix expansion of (~\ref{block1}), the eigenvalue of interest
$E_0$ is now shielded from the rest of the spectrum by $E_1$. The solution
of the eigenproblem for ${\tilde H}_0$ generally leads to a non-polynomial
dependence of ${\tilde E}_0$ as a function of $g$. For instance, the first 
order corrected ground state energy for the matrix method is
given by the expression


\begin{eqnarray}
\nonumber{\tilde E}_0 = \frac{1}{2} (E_0+gV_{00}+E_1+gV_{11})- \\
\frac{1}{2}\sqrt{E_1+gV_{11}-E_0-gV_{00}+4g^2V_{01}^2}.
\label{blockener}
\end{eqnarray}


It is clear that if $g/(E_1-E_0) \agt 1/2$, the expansion
(~\ref{expansion1}) will diverge while the expression of ${\tilde E}_0$ in 
 (~\ref{blockener}) is well defined, i.e. non-trivial effects are
already included into ${\tilde E}_0$ at the first order of the matrix expansion.

Let us now study a few models in order to illustrate the 
difference between the matrix method and the
simple polynomial expansion. It is interesting to first study the Mathieu
equation which arises, for instance, after separation of variables of
Laplace's equation in elliptic cylindrical coordinates. This is 
because this equation was actually used by Bloch \cite{bloch} as a test 
for the polynomial expansion. The Mathieu equation is

\begin{eqnarray}
(-\frac{d^2}{d\theta^2}+g\cos^2 \theta)\phi(\theta)=E\phi(\theta).
\label{mathieu}
\end{eqnarray}

This equation may be studied by perturbation theory with $d^2/d\theta^2$
as the $H_0$ and $g \cos^2 \theta$ as the perturbation. The eigenvalues $E_n$ 
and eigenfunctions $\phi_n$ of the free part which are even functions
with period $\pi$ are $E_n=n^2$  and $\phi_n=1/\sqrt{\pi}\cos n\theta$ where
$n=0,2,4...$. The correction to the groundstate energy obtained from the 
original Bloch method up to the fourth order is

\begin{eqnarray}
{\tilde E}_0^{(4)}= \frac{g}{2}-\frac{1}{2}(\frac{g}{4})^2+\frac{7}{128}(\frac{g}{4})^4+O(g^5).
\label{eperturb}
\end{eqnarray}

\begin{table}
\begin{ruledtabular}
\begin{tabular}{cccccc}

 $g$ & Bloch &  $m=2$&  $m=4$&  $m=8$& exact \\
\hline

 0.5 &  0.242201 &  0.242203 & 0.242201 &  0.242201 & 0.242201 \\
 1   &  0.468964 &  0.468910 & 0.468961 &  0.468961 & 0.468961 \\
 2   &  0.878418 &  0.878680 & 0.878234 &  0.878234 & 0.878234 \\
 4   &  1.554688 &  1.550510 & 1.544861 &  1.544861 & 1.544861 \\
 8   &  2.875000 &  2.535818 & 2.486044 &  2.486044 & 2.486044 \\
 16  & 14.000000 &  4.000000 & 3.719515 &  3.719481 & 3.719481 \\

\end{tabular}
\end{ruledtabular}
\caption{  Fourth order simple expansion from (~\ref{eperturb}) (Bloch) versus
first order matrix expansion for $m=2,4,8$ states
kept in for the  lowest state of the Mathieu equation.}
\label{bloch-block}
\end{table}

 Table~\ref{bloch-block} compares the correction obtained from 
(~\ref{eperturb}) with the matrix perturbation theory and the exact result.
When $g \alt 2$, the condition $g/(E_1-E_0) < 1/2$ is fulfilled, both the
first order matrix perturbation result even for a small number of states
$m$, and the fourth order perturbation estimate of (~\ref{eperturb}), 
shown in Table~\ref{bloch-block}, agree quite well with the exact solution. 
One may note that the first order simple perturbation estimate is $0.250000$.
At this level, the first order simple perturbation and first order
matrix perturbation are already different as explained above. When $g \agt 4$, 
$g/(E_1-E_0) \agt 1$ the simple perturbation series diverges and 
(~\ref{eperturb}) cannot be used to compute the correction to the ground 
state energy. The difference between the simple perturbation and the
exact result increases with increasing $g$. In contrast, 
the matrix perturbation method leads to good results up to $g=16$ even if a
small number of states is kept in the construction of $P_0$. The agreement
with the exact result extends to more than the sixth digit when eight or
more states are kept. The matrix perturbation estimate can be improved
for a fixed number of states by increasing the order of the matrix series.
For instance in Table~\ref{bloch-block2}, when a second order term is 
included for $m=2$, the agreement with the exact result  is better.
In the Mathieu equation, the rapid convergence of the matrix method
is due to the fact that the energy separation between consecutive
eigenvalues is roughly $(n+2)^2-n^2$. Thus, the condition 
$g/(E_{n_c+1}-E_0) < 1/2$ can easily be satisfied.

\begin{table}
\begin{ruledtabular}
\begin{tabular}{cccc}

 $g$ & $m=2$(1)& $m=2$(2) & exact \\
\hline

 0.5 &  0.242203 & 0.242201 & 0.242201 \\
 1   &  0.468910 & 0.468961 & 0.468961 \\
 2   &  0.878680 & 0.878231 & 0.878234 \\
 4   &  1.550510 & 1.544708 & 1.544861 \\
 8   &  2.535818 & 2.481531 & 2.486044 \\
 16  &  4.000000 & 3.647650 & 3.719481 \\

\end{tabular}
\end{ruledtabular}
\caption{ First order(1) versus second order (2) 2-state matrix
expansion for the lowest state of the Mathieu equation.}
\label{bloch-block2}
\end{table}

 Let us now consider the case of an harmonic oscillator with a quartic 
perturbation which does not present this advantage. The Hamiltonian is

\begin{eqnarray}
H_{osc}= \frac{1}{2}(p^2+q^2)+gq^4.
\end{eqnarray}
 
In the Dirac notations, $H_{osc}$ becomes 

\begin{eqnarray}
\nonumber H_{osc}= a^{\dagger}a+\frac{1}{2}+\frac{g}{4}(a^4+a^{\dagger 4}+ 
4a^{\dagger 3}a + 4a^{\dagger }a^3 + \\
6a^{\dagger 2}a^2+ 6a^2 + 6a^{\dagger 2}+ 12a^{\dagger}a+3).
\end{eqnarray}

The unperturbed energies are now $E_n=0$, $1$, $2$,...This model has
widely been used to test different perturbative approaches. It is
now well established that the simple Brillouin-Wigner series is divergent
for this model. One has to resort to special resummation procedures 
in order to obtain convergence. Table~\ref{oscillator} shows that a simple 
first order matrix approach can accurately reproduce the exact result up to six 
digits for a modest number of states kept. But as expected, since the 
energy separation is $1$, a larger $m$ than in the Mathieu equation needs 
to be used in order to achieve the same accuracy.

\begin{table}
\begin{ruledtabular}
\begin{tabular}{cccccc}

 $g$ & $m=4$ &  $m=8$&  $m=16$&  $m=32$& exact \\
\hline

 0.1 &  0.559564 &  0.559165 & 0.559146 &  0.559146 & 0.559146 \\
 0.3 &  0.640354 &  0.638539 & 0.637992 &  0.637992 & 0.639992 \\
 0.5 &  0.706301 &  0.697454 & 0.696178 &  0.696176 & 0.696176 \\
 1.0 &  0.855087 &  0.805870 & 0.803837 &  0.803771 & 0.803771 \\
 2.0 &  1.137219 &  0.956286 & 0.952468 &  0.951571 & 0.951568 \\

\end{tabular}
\end{ruledtabular}
\caption{ First order matrix expansion results for $m=4,8,16, 32$ states
kept compared to the exact result of reference \cite{janke} for the  
lowest state of anharmonic oscillator.}
\label{oscillator}
\end{table}


 Let us now consider a non-trivial model, antiferromagnetic (AF) Heisenberg
chains weakly coupled by a ferromagnetic transverse exchange $J_{\perp}<0$. 
 In this problem, $H_0=H_{\parallel}$ is an array of decoupled AF chains.
 When the conventional random phase approximation (RPA) is applied to this
problem, one finds that the spin suceptibility diverges at low temperatures
for any small $J_{\perp}$. This indicates that the ground state is ordered
as soon as $ J_{\perp} \neq 0$. But starting from the disordered chain,
the ordered regime cannot be reached by RPA. It is necessary to turn to
special procedures such as the chain-mean field approach 
\cite{scalapino,schulz} in which the
existence of long-range order is assumed. It will now be shown below that the
matrix Kato-Bloch method can reach the ordered regime without assuming
long-range order {\it a priori}. 

\begin{figure}
\includegraphics[width=3. in]{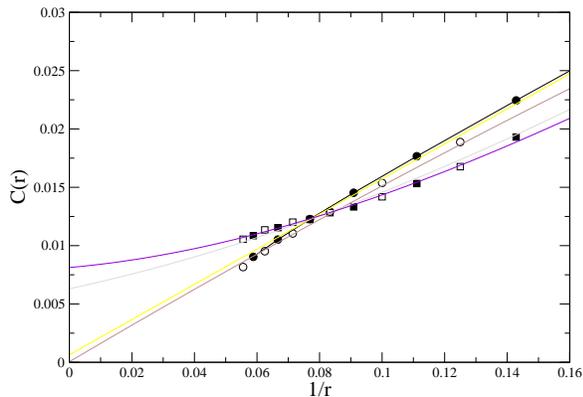}

\caption{The ground state correlation function
$C(r)={\bar C}_{\parallel}(25,25,r)$ for the $48 \times 49$
lattice for $J_{\perp}=0$  (circles) and
$J_{\perp}=-0.16$  (squares). The filled and open symbols 
correspond to odd and even distances respectively.}
\label{spincorl}
\end{figure}

 The exact spectrum of a single AF chain is known from the Bethe
ansatz, but eigenfunctions are not easily accessible. Thus,
the density-matrix renormalization group (DMRG) method \cite{white} 
will be used to compute an approximate spectrum $\epsilon_n, |\phi_n \rangle$ 
of a single chain.  A preliminary account of the first order of this approach 
\cite{moukouri_TSDMRG} as well as an extensive comparison with the 
Quantum Monte Carlo method was presented elsewhere \cite{moukouri_TSDMRG2}. 
By expressing the Hamiltonian on the basis generated 
by the tensor product of the states of different chains one obtains, 
up to the second order, the effective one-dimensional Hamiltonian,

\begin{eqnarray}
\nonumber \tilde{H} \approx \sum_{[n]} E_{\parallel [n]} 
|\Phi_{\parallel[n]}\rangle \langle \Phi_{\parallel [n]}| +
 J_{\perp} \sum_{l} {\bf \tilde{S}}_{l} {\bf \tilde{S}}_{l+1}-\\
\frac{J_{\perp}^2}{2} \sum_{l} {\bf S}_{l}^{(2)}{\bf S}_{l+1}^{(2)}
+...
\label{efhamil}
\end{eqnarray}

\noindent where the chain-spin operators on the chain $l$ are 
${\bf \tilde{S}}_{l}=({\bf \tilde{S}}_{1l},
 {\bf \tilde{S}}_{2l}, ...{\bf \tilde{S}}_{Ll})$ and 
${\bf S}_{il}^{(2)}=({\bf \tilde{S}}_{1l}^{(2)},
 {\bf \tilde{S}}_{2l}^{(2)}, ...{\bf \tilde{S}}_{Ll}^{(2)})$, $L$ is the 
chain length. The matrix elements of the first and second order local 
spin operators are respectively

\begin{eqnarray}
\nonumber {\bf \tilde{S}}_{i,l}^{n_l,m_l}=\langle \phi_{n_l}|{\bf S}_{i,l}|\phi_{m_l}\rangle,
\\
{\bf S}_{il}^{(2)n_ln'_l}=\sum_{m_l}\frac{{\tilde S}_{il}^{n_lm_l}
{\tilde S}_{il}^{m_ln'_l}}{\sqrt{\epsilon_{m_l}-\epsilon_{0_l}}}.
\end{eqnarray}

The spectrum of the effective Hamiltonian (~\ref{efhamil}) which 
is one-dimensional is also obtained by applying the DMRG. It is to 
be emphasized that the use of the DMRG here stems from the one-dimensionality 
of the effective problem. But in general, it is not necessary to apply this 
technique. $m=80$ states were kept which leads to 
$J_{\perp}/(E_{n_c+1}-E_0) \approx 0.26$ for $J_{\perp}=-0.16$, which means
the Rellich theorem is satisfied. 
Fig.(~\ref{spincorl}) shows the spin-spin correlation function
parallel to the chains, ${\bar C}_{\parallel}(25,25,r)=\langle{\bf S(0)S(r)}\rangle/3$ 
for the middle chain $l=25$, the origin taken on site $i=25$ in 
the $48 \times 49$ lattice, for the cases $J_{\perp}=0$ and $J_{\perp}=-0.16$.
It is clearly seen that, as expected, the former extrapolate to zero. The
same behavior is observed on $32 \times 33$ and $64 \times 65$ lattices.
 But when $J_{\perp}=-0.16$ the extrapolated value is finite. This
indicates the presence of long-range order.  The same behavior is observed 
on $32 \times 33$ and $64 \times 65$ lattices (not shown). The effect of
the other chains is to create an effective magnetic field on the middle chain
which leads to a finite order parameter. This result thus justifies the
assumption behind chain mean-field theory approaches \cite{scalapino, schulz}.

In summary, using a somewhat loose interpretation of the Rellich theorem,
a convergent matrix perturbation approach has been proposed.
 This new method, which treats both the ordered and disordered regimes
in a controlled way,  opens new possibilities in the study of phase 
transitions and strongly correlated electron in condensed-matter systems.
 The new method may also be useful in handling infrared divergences in
light-front Hamiltonians in quantum chromodynamics \cite{wilson}.

\begin{acknowledgments}
I wish to thank J.W. Allen for helpful discussions and P. McRobbie
for reading the manuscript.
\end{acknowledgments}



\end{document}